\documentclass[conference]{IEEEtran}
\IEEEoverridecommandlockouts

\usepackage{cite}
\usepackage{graphicx}
\usepackage{amsmath,amssymb,amsfonts}
\usepackage{textcomp}
\def\BibTeX{{\rm B\kern-.05em{\sc i\kern-.025em b}\kern-.08em
    T\kern-.1667em\lower.7ex\hbox{E}\kern-.125emX}}
\usepackage{pifont} 
\usepackage{xcolor} 
\usepackage{bm}
\usepackage{array}
\usepackage{balance}

\usepackage{colortbl}
\usepackage{xspace}
\usepackage[framemethod=TikZ]{mdframed}
\usepackage{textcomp}
\usepackage{stfloats}
\usepackage{multirow}
\usepackage{url}
\usepackage{verbatim}
\usepackage[most]{tcolorbox}
\tcbuselibrary{breakable}
\usepackage{caption}
\usepackage{multirow}
\usepackage{booktabs}
\usepackage{calc}
\usepackage{bbding}
\usepackage{threeparttable}
\usepackage{algorithm}
\usepackage{algpseudocode}
\usepackage{subfigure}
\usepackage{float}

\newcommand{\rqone}{RQ1: What is the effectiveness of different retrievers?\xspace}

\newcommand{\rqtwo}{RQ2: What is the efficiency of different retrievers?\xspace}

\newcommand{\rqthree}{RQ3: What is the effectiveness-efficiency trade-off of different retrievers?\xspace} 

\newcommand{\rqfour}{RQ4: What is the impact of the hyper-parameters of retrievers on their efficiency and effectiveness?\xspace}

\newcommand{\rqfive}{RQ5: What is the impact of the number of shots?\xspace}

\begin{document}


\title{Evaluating the Effectiveness and Efficiency of Demonstration Retrievers in RAG for Coding Tasks}

\author{\IEEEauthorblockN{Pengfei He, Shaowei Wang, Shaiful Chowdhury}
\IEEEauthorblockA{\textit{University of Manitoba} \\
Winnipeg, Canada\\
hep2@myumanitoba.ca, \{shaowei.wang, shaiful.chowdhury\}@umanitoba.ca
}

\and
\IEEEauthorblockN{Tse-Hsun Chen}
\IEEEauthorblockA{
\textit{Concordia University}\\
Montreal, Canada \\
peterc@encs.concordia.ca}
}

\maketitle

\begin{abstract}
Retrieval-Augmented Generation (RAG) enhances Large Language Models (LLMs) by integrating external knowledge bases, achieving state-of-the-art results in various coding tasks. The core of RAG is retrieving demonstration examples, which is essential to balance effectiveness (generation quality) and efficiency (retrieval time) for optimal performance. However, the high-dimensional nature of code representations and large knowledge bases often create efficiency bottlenecks, which are overlooked in previous research. This paper systematically evaluates the efficiency-effectiveness trade-off of retrievers across three coding tasks: Program Synthesis, Commit Message Generation, and Assertion Generation. We examined six retrievers: two sparse (BM25 and BM25L) and four dense retrievers, including one exhaustive dense retriever (SBERT's Semantic Search) and three approximate dense retrievers (ANNOY, LSH, and HNSW). Our findings show that while BM25 excels in effectiveness, it suffers in efficiency as the knowledge base grows beyond $10^3$ entries. In large-scale retrieval, efficiency differences become more pronounced, with approximate dense retrievers offering the greatest gains.
For instance, in Commit Generation task, HNSW achieves a 44x speed up, while only with a 1.74\% drop in RougeL compared with BM25. Our results also demonstrate that increasing the number of demonstrations in the prompt does not consistently enhance effectiveness. Instead, it can increase latency and lead to incorrect outputs. Our
findings provide valuable insights for practitioners aiming to build efficient and effective RAG systems for coding tasks.
\end{abstract}


\section{Introduction}

Large language models (LLMs) have achieved remarkable success in natural language processing, but they still encounter significant limitations in the domain of knowledge-intensive tasks. In particular, LLMs are susceptible to ``hallucinations'' when confronted with queries that exceed the scope of their training data or necessitate the utilization of current information \cite{codehalu}. Retrieval-Augmented Generation (RAG) is the leading technique for improving LLMs by providing demonstrations from knowledge bases. By referencing external knowledge sources, RAG effectively mitigates the problem of generating factual inconsistency content and facilitates the continuous updating of knowledge~\cite{ragsurvey}.


Recently, the application of RAG has achieved promising results for various code-specific tasks such as Code Search~\cite{allyouneed}, Program Synthesis~\cite{whatmakes,docprompting}, and Assertion Generation~\cite{UBC}. In these tasks, LLM learns from contextual prompts consisting of task descriptions, queries, and additional demonstration examples without the need to fine-tune the model parameters. The retrieved demonstrations are typically used as the context to assist the pre-trained LLM in comprehending the task and regulating the generation behavior, which usually has a significant impact on the quality of the generated output. Therefore, it is important to retrieve appropriate demonstrations from a vast knowledge base.

The retriever plays a core role in retrieving relevant demonstrations from the external knowledge base and significantly affects the performance of RAG~\cite{sawarkar2024blended,salemi2024evaluating}. Retrievers are typically classified into sparse and dense retrievers based on representation methods. Sparse retrievers operate at the token level, while dense retrievers operate at the level of latent semantics. The most widely-used sparse retriever, BM25~\cite{rankbm25}, ranks demonstrations based on term frequency (TF) and inverse document frequency (IDF) of the query. Dense retrievers perform retrieval by encoding the query and demonstrations into dense embedding representations and scoring each demonstration by its similarity with the query embedding. For instance, the popular RAG system Llamaindex~\cite{llamaindex} supports both BM25 and custom embedding encoders in RAG workflow. Similarly, recent approaches for coding tasks, such as the work by Nashid et al.~\cite{UBC}, Zhou et al. ~\cite{docprompting}, and Gao et al.~\cite{whatmakes} use both BM25 and BERT-based encoder (e.g., SBERT \cite{sentencebert} and Unixcoder \cite{unixcoder}) to select code samples from their demonstration pools. In these studies, RAG demonstrated superior performance compared to task-specific models across various code-related tasks, highlighting the critical role of retrieving semantically relevant demonstrations.

When deploying RAG systems, it is essential to achieve a balance between ``effectiveness'' (quality of the generated results) and ``efficiency'' (runtime of demonstrations retrieval), typically when the knowledge base is large. However, existing studies in the RAG for code-specific tasks lack attention to efficiency and have some limitations. \textbf{(1) Insufficient trade-off consideration:} When RAG is implemented on a large scale, significant retrieval time gaps are observed between different retrievers. For instance, in our preliminary evaluation of the Assertion Generation Task~\cite{atlas}, involving 18k test samples and a knowledge base of 150k entries, BM25 required approximately 30 hours to complete the retrieval process, while SBERT accomplished the same task in just 1257 seconds. Despite BM25's significant efficiency disadvantage, it achieved only a marginal 1\% increase in the Exact Match rate. This highlights the trade-offs between efficiency and effectiveness in large-scale data processing scenarios. \textbf{(2) High complexity of Exhaustive Nearest Neighbor Search:} Exhaustive dense retrievers depend on performing an exhaustive nearest neighbor search after encoding, requiring a linear scan across the entire knowledge base to find the most relevant demonstrations. For instance, SBERT's Semantic Search, which utilizes 768-dimensional SBERT embeddings, results in a computational complexity of $O(768n)$ when searching through a knowledge base with $n$ entries. This high computational overhead makes the approach inefficient, severely limiting the scalability of RAG systems, particularly when working with large-scale code knowledge bases.

To address these limitations, we conducted a comprehensive study to assess the effectiveness and efficiency of various retrievers in coding tasks utilizing RAG. Specifically, we evaluated six retrievers across three tasks: Program Synthesis, Commit Message Generation, and Assertion Generation. For sparse retrievers, we used BM25 and its variant, BM25L~\cite{bm25L}. For dense retrievers, we first encoded queries and demonstrations of knowledge base as embedding vectors using SBERT~\cite{sentencebert}. We then employed two types of search strategies, Exhaustive and Approximate Nearest Neighbor (ANN) search, to find the demonstration in a knowledge base that is closest to a query data point. SBERT’s Semantic Search (SBERT\_SS) is a widely-used exhaustive dense retriever \cite{whatmakes,longllmlingua,UBC}, performing a linear scan by calculating cosine similarity between the query and all demonstrations in the knowledge base. In contrast, we introduced three ANN-based approaches for approximate dense retrieval: Approximate Nearest Neighbors Oh Yeah (ANNOY)\cite{annoy}, Locality-Sensitive Hashing (LSH)\cite{LSH}, and Hierarchical Navigable Small World (HNSW)~\cite{hnsworg}. Unlike exhaustive retrieval, these approximate approaches retrieve a subset of relevant demonstrations using approximation algorithms, significantly speeding up the retrieval process by limiting the search to a smaller subset rather than the entire knowledge base.



We find that while the sparse retriever BM25 excels in effectiveness, its efficiency becomes the worst when the knowledge base exceeds $10^3$ entries. For larger-scale retrieval, approximate dense retrievers offer significant efficiency gains, achieving a better trade-off between efficiency and effectiveness. For example, in the Commit Generation task, HNSW delivers a 44x speedup with only a 1.74\% drop in RougeL compared to BM25. However, a high degree of approximation, while improving efficiency, may reduce retrieval precision and impact overall effectiveness. Additionally, we observe that adding more demonstrations to the prompt does not necessarily enhance RAG effectiveness, often increasing response latency and leading to incorrect outputs. These findings provide important insights for practitioners aiming to build efficient and effective RAG systems for coding tasks.

In summary, the contributions of our paper include:
\begin{itemize}
    \item We conducted the first systematic study to assess both the efficiency and effectiveness of various retrievers in RAG for coding tasks with extensive experiments.
    \item Our findings provide actionable insights into the selection of retrievers for practitioners aiming to implement RAG for code-specific tasks. \textbf{Yeas 1:} Using approximate approaches (e.g., HNSW and LSH) for RAG can significantly speed up retrieval with negligible degradation in generation quality for large knowledge bases (e.g., exceeds $10^3$ entries). \textbf{Yeas 2:} Using sparse retrievers such as BM25 which excels in effectiveness when the knowledge base is small. \textbf{Nays1: } Be cautious! More demonstrations do not guarantee a better generation.
    \item  We release our replication package~\cite{ANN4RAG} to facilitate future research. 

\end{itemize}

\section{Background}
\begin{figure*}[tp]
\centering

\includegraphics[width=1\linewidth]{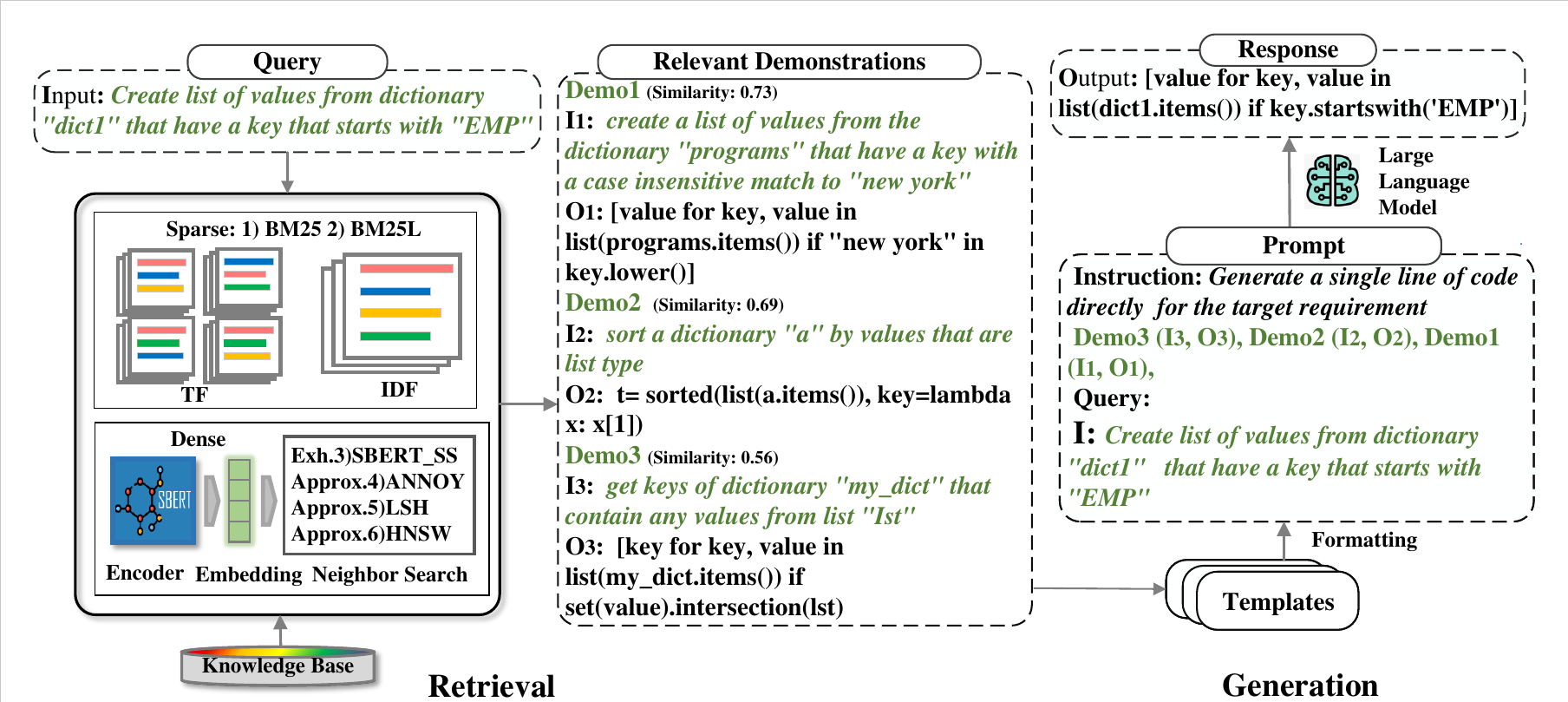}
\caption{The general framework of RAG for coding tasks.}
\label{fig:framework}
\end{figure*}



\subsection{Retrieval Augmentation Generation}
RAG is the leading prompt engineering technique designed to enhance the output of LLMs by integrating external knowledge through a two-phase process: \textbf{Retrieval} and \textbf{Generation}. The primary purpose of RAG is to provide LLMs with relevant demonstrations as context, thereby improving its ability to generate accurate and context-relevant outputs~\cite{ragsurvey}.

In the \textbf{Retrieval} phase, the objective is to retrieve demonstrations that are highly relevant to the query for the subsequent \textbf{Generation} phase. For example, as depicted in Fig.~\ref{fig:framework} in the context of a Program Synthesis task, the process begins with a query input $I_{query}$ asking how to implement a specific function. The retriever returns Top $K$ relevant demonstrations, forming the set $D_K$,  where each demonstration $demo_i$ is composed of similar requirements $I_i$ and corresponding code snippets $O_i$ as represented by $D_K = \{demo1: (I_1, O_1), demo2: (I_2, O_2), \dots, demoK: (I_{K}, O_{K})\}$.
                        
In the \textbf{Generation} phase, the query and the demonstrations are combined to create a retrieval-augmented prompt using a predefined template, denoted as $Template(D_K, I_{query})$. The $K$ demonstrations in $D_K$ can be arranged in different ways, such as by similarity to the query, in reverse similarity order~\cite{whatmakes}, or reordered to avoid the ``lost in the middle'' issue~\cite{lost}, where demonstrations at both ends tend to have more influence than those in the middle. Following the recommended setup in ICL4code~\cite{whatmakes}, we adopt a similarity-based order in this study, positioning the most relevant demonstration closest to the query. This retrieval-enhanced prompt is then fed into the LLM to generate the final output.

\subsection{Retrieval Approach}
Typically, there are two families of retrieval approaches, namely \textbf{Sparse} and \textbf{Dense}.

\noindent\textbf{Sparse} retrievers rank demonstrations by evaluating the frequency of terms within a demonstration (TF) and their rarity across the entire knowledge base (IDF). Specifically, for each term in the query, a TF/IDF score is calculated for every demonstration in the knowledge base. Demonstrations that have higher cumulative TF/IDF scores are considered more relevant. BM25\cite{bm25} and its variants (e.g., BM25L\cite{bm25L}, BM25+\cite{bm25compare}) improve upon TF/IDF by normalizing the TF and adjusting for the saturation effect. This results in a more nuanced ranking that better reflects the relevance of demonstrations to the query.

\noindent\textbf{Dense} retrievers leverage high-dimensional numeric representations to encode demonstrations and queries, with cosine similarity as a common metric to measure their relevance. For instance, using SBERT~\cite{sentencebert} as the encoder, SBERT's Semantic Search implements a straightforward and widely-used strategy to perform a linear scan of the entire knowledge base $D$, commonly referred to as an \textbf{exhaustive} dense retriever~\cite{UBC,whatmakes,docprompting}, as expressed by:
\begin{equation}
D_K=\arg \max _{D_K \subset D,|D_K|=K} \sum_{demo_i \in D} \operatorname{Sim}\left(I_{query}, demo_i\right) 
\end{equation}
This process incurs a computational complexity of $O(mn)$, where $m$ is the dimensionality of the embeddings and  $n$ is the size of the knowledge base. As the dataset grows, this complexity becomes impractical due to \emph{the curse of dimensionality}, making exhaustive searches increasingly inefficient for large-scale retrieval tasks~\cite{powell2007approximate}. To enable efficient RAG, it is necessary to reduce the computational complexity of the retrieval process. Therefore, we also explore \textbf{approximate} dense retrievers~\cite{powell2007approximate}, which leverage ANN techniques that have not been thoroughly investigated in previous studies on coding tasks. Approximate dense retrievers build indexes on $D$.  When a query is given, it retrieves a subset $A$ of $D$ by indexes and then finds relevant demonstrations in $A$ rather than scanning the entire knowledge base. In this way, approximate dense retrievers can get an approximation of true similar demonstrations through reducing the search neighbors as: 
\begin{equation}
D_K=\arg \max _{D_K \subset A,|D_K|=K} \sum_{demo_i \in A \subset D} \operatorname{Sim}\left(I_{query}, demo_i\right) 
\end{equation}

Various ANN techniques exist to build indexes, including tree-based (e.g., KDTree\cite{kdtree}, ANNOY\cite{annoy}), graph-based (e.g., HNSW \cite{faiss,hnsworg}), and hash-based approaches (e.g., LSH \cite{faiss,LSH}). For instance, ANNOY creates a tree structure that partitions the data space, enabling early termination of the search process once a sufficiently close neighbor is identified. HNSW constructs a graph in which each vertex represents a data point linked to its true nearest neighbors, enabling efficient traversal to locate approximate nearest neighbors. Hash-based approaches, such as LSH, optimize search efficiency by reducing the dimensionality of data points using specialized hashing functions.

\subsection{Related work}

RAG has been increasingly adopted in coding tasks to improve the quality of output generated by LLMs~\cite{allyouneed,UBC,docprompting,whatmakes}. Chen et al.\cite{allyouneed} explored a RAG framework for Code Suggestion, employing four retrieval strategies to identify similar code demonstrations. They observed that integrating retrieved demonstrations significantly improved outputs of LLMs in Code Suggestion tasks.  Similarly, Zhou et al.~\cite{docprompting} proposed a text-to-code generation framework that utilizes retrieved API documentation, ensuring LLMs stay up-to-date with the latest API changes. Nashid et al.~\cite{UBC} demonstrated the effectiveness of RAG approaches in more complex tasks, such as Program Repair and automatic Assertion Generation, achieving notable improvements without requiring task-specific fine-tuning. In a systematic study, Gao et al. ~\cite{whatmakes} analyzed the influence of retrieval demonstrations on various coding tasks, such as Code Summarization and Bug Fixing. They investigated optimal strategies for selecting, ordering, and determining the quantity of demonstrations.

Among these studies, sparse retrievers like BM25 and dense exhaustive retrievers utilizing encoders have become standard approaches in RAG workflows. However, no comprehensive studies have evaluated the effectiveness-efficiency trade-off between different retrievers. This trade-off is critical but often overlooked, as it significantly impacts both generation quality and runtime. Additionally, we are the first to investigate approximate dense retrievers in RAG for coding tasks using Approximate Nearest Neighbor search (ANN) \cite{annbenchmarks}, a range of approaches widely studied in high-dimensional fields such as computer vision and recommendation systems~\cite{graphann}, but unexplored in RAG domain.




\section{Experimental Design}\label{sec:experimentalsetting}
In this section, we present our research questions (RQs),
studied dataset, retrievers, used prompt templates, implementation details, and our approach to RQs. We conduct all our experiments following the framework presented in Fig.~\ref{fig:framework}.

\subsection{Research Question}
We aim to answer the following research question:
\begin{itemize}
    \item \textbf{\rqone}
    \item \textbf{\rqtwo}
    \item \textbf{\rqthree}
    \item \textbf{\rqfour}
    \item \textbf{\rqfive}
\end{itemize}


In RQ1, we aim to evaluate the effectiveness of various retrievers by comparing how well they retrieve relevant demonstrations that enhance output quality. RQ2 focuses on the efficiency of these retrievers, assessing retrieval time and scalability as the knowledge base grows. RQ3 investigates the trade-off between effectiveness and efficiency, identifying retrievers that balance high-quality output with minimal retrieval time. RQ4 explores approximate dense retrievers, analyzing how parameter adjustments impact both retrieval time and generation quality. Finally, RQ5 examines the effect of the number of demonstrations, determining whether increasing the number of shots consistently improves effectiveness.

\subsection{Dataset and Metric}\label{sec:dataset}
To answer our RQs, we conducted our study on three popular coding tasks: Program Synthesis \cite{conala,whatmakes,docprompting}, Commit Message Generation \cite{commitnngen}, and Assertion Generation \cite{atlas}.

Program Synthesis focuses on generating code snippets from natural language requirements. We employed the CoNaLa dataset~\cite{conala}, which contains 2,889 NL-code pairs from Stack Overflow (Python). Following prior research~\cite{whatmakes}, we used four evaluation metrics: Code Bleu~\cite{CodeBleu}, Syntax Match, Dataflow Match, and N-gram Match rate. These metrics capture both surface- and semantic- level correctness, with Code Bleu extending traditional BLEU for code features. 

Commit Message Generation involves creating descriptive messages to accompany code changes during commits. For this, we utilized the NNgen dataset~\cite{commitnngen}, which consists of diff-commit pairs from Git repositories. The diff serves as the input, representing the changes between two repository versions, while the commit serves as the output, summarizing the change. We evaluated the task using RougeL, Meteror,  Rouge1, and Rouge2, following previous work~\cite{commitnngen}. RougeL measures the longest common subsequence between the generated and reference messages, while Meteor combines unigram precision and recall. Rouge1 and Rouge2 assess unigram and bigram overlaps, respectively.

Assertion Generation automates the creation of assertion statements to verify the correctness of a focal method (i.e., a method under
test). We used the Atlas dataset~\cite{UBC}, which contains pairs of focal and test methods to generate assertion statements. Following \cite{UBC}, we employed four evaluation metrics: Exact Match, Plausible Match, Longest Common Subsequence, and Edit Distance. Exact Match rate measures the proportion of predictions that exactly match the ground truth, while Plausible Match accounts for semantically equivalent assertions (e.g., treating $assertTrue(value == 0)$ and $assertEquals(value, 0)$ as identical). Longest Common Subsequence evaluates the overlap between predicted and expected assertions, and Edit Distance calculates the number of edits needed to match the prediction to the ground truth.

For RQ1, we report the effectiveness in terms of all introduced evaluation metrics. However, given the space constraints and the correlation among metrics, in RQ2 to RQ5, we primarily focus on \textbf{Code Bleu}, \textbf{RougeL}, and \textbf{Exact Match} for Program Synthesis, Commit Generation, and Assertion Generation, respectively.

The statistics for these three tasks are presented in Table~\ref{dataset}. We used the original dataset partitions, with the training set serving as the knowledge base and the test set for evaluation. Our dataset consists of tasks with different size of knowledge base for retrieval.

\begin{table}[htbp]
\caption{Statistics of our studied datasets.} 
\label{dataset}
\centering
\begin{tabular}{l|ccc}
\hline
\textbf{Task}        & \textbf{Dataset} & \textbf{Knowledge Base} & \textbf{Test} \\ \hline
Program Synthesis    & CoNaLa\cite{whatmakes,conala}            & 2,300                        & 477                \\
Commit Generation    & NNGen\cite{commitnngen}            & 22,112                        & 2,521               \\
Assertion Generation & Atlas\cite{UBC}            & 150,523                     & 18,027             \\ \hline
\end{tabular}
\end{table}

\subsection{Studied Retrievers}\label{sec:retriever}

To ensure reliable insights, we selected representative retrievers from both sparse and dense retrieval families for their broad adoption in various studies \cite{UBC,whatmakes,longllmlingua}. For sparse retrievers, we chose two retrievers: BM25~\cite{bm25} and BM25L~\cite{bm25L}. For dense retrievers, we selected SBERT\_SS~\cite{sentencebert,UBC,whatmakes} as the non-approximate dense retriever. For approximate dense retrievers, we selected three ANN approaches based on different underlying structures: ANNOY~\cite{annoy,annbenchmarks} for tree-based approaches, LSH~\cite{LSH} for hash-based approaches, and HNSW~\cite{hnsworg} for graph-based approaches.

BM25\cite{bm25} is an advanced ranking function used in information retrieval to assess the relevance of documents to a search query. Building on the conventional TF-IDF model, BM25 improves by incorporating document length normalization and a non-linear term frequency function. We utilize the rank-bm25 implementation \cite{rankbm25}, which has also been employed in previous research \cite{llamaindex, UBC, longllmlingua, docprompting,whatmakes}.

BM25L\cite{bm25L} is a variant of BM25 that modifies the length normalization factor to enhance retrieval performance for longer documents. This adaptation addresses the bias toward shorter documents present in the original BM25 model. Like BM25, we also use the rank-bm25 implementation.

SBERT\_SS \cite{SBERTlink} is a built-in method of SBERT that computes the cosine similarity between the query vector and all demonstration vectors in the knowledge base. SBERT\_SS has been widely used in previous works \cite{docprompting, UBC, whatmakes,lingua2}.

ANNOY \cite{annoy} constructs a forest of binary trees using random projections. Each non-leaf node acts as a hyperplane, splitting the space into two partitions. During a search, ANNOY traverses these trees, comparing the query point to points in each leaf node and updating the current nearest data point. The hyper-parameter \textbf{Search\_K} (the number of nodes inspected) controls the trade-off between precision and efficiency. ANNOY is a competitive ANN approach in the ANN benchmark\cite{annbenchmarks}.

LSH \cite{LSH} reduces data dimensionality while preserving local distances. For a binary flat index, LSH converts data points into binary vectors, increasing the likelihood that similar data points are hashed into the same bucket. The hyper-parameter \textbf{Nbit} (the number of bits) controls the trade-off between precision and efficiency. We use the faiss\cite{faiss} implementation, which is also utilized in the ANN benchmark\cite{annbenchmarks}.

HNSW \cite{hnsworg} organizes data points into hierarchical layers and leverages small-world network properties. The search begins at the top layer and moves downward, with each layer refining the search by following connections to nodes closer to the query. The hyper-parameter \textbf{M} (the number of neighbors in the graph) controls the trade-off between precision and efficiency. We use the faiss implementation for HNSW as well.

\begin{table*}[htbp]
\caption{Prompt templates for the three studied tasks.}
\label{temp}

\centering
\begin{tabular}{l|l|l|l}
\hline
Task     & \multicolumn{1}{c|}{Program Synthesis}                                                               & \multicolumn{1}{c|}{Commit Generation}                                                                                                                                        & \multicolumn{1}{c}{Assertion Generation}                                                                                                                                     \\ \hline
Templates & \begin{tabular}[c]{@{}l@{}}Generate only single line code\\ for the target requirement directly,\\ and nothing else.\\
\#\#\# requirement: {[}requirement{]}\\ \#\#\# code: {[}code{]}\\ end\_of\_demo\\ ...\\ \#\#\# target requirement: {[}requirement{]}\\\#\#\# code: {[}code{]}\end{tabular} & \begin{tabular}[c]{@{}l@{}}Generate commit message for \\
query diff: {[diff]}\\
commit message: [commit message]\\
only return the [commit message], \\and nothing else.\\
Make sure your output is for query diff\\ and concise.\\
Here are some reference demonstrations:\\
diff: [diff],\\commit message: [commit message]
\\...
\end{tabular} & \begin{tabular}[c]
{@{}l@{}}Generate only assertions based on DEMO,\\ and nothing else.\\\#\#\# METHOD\_UNDER\_TEST: {[}focal method{]}\\ \#\#\# UNIT\_TEST\: {[}test method{]}\\ \#\#\# generate assertion:
{[}assertion{]}\\end\_of\_demo\\ ...\\\#\#\# METHOD\_UNDER\_TEST: {[}focal method{]}\\ \#\#\# UNIT\_TEST\: {[}test method{]}\\ \#\#\# generate assertion:
{[}assertion{]}

 \end{tabular} \\ \hline
\end{tabular}
\end{table*}

\subsection{Prompt Templates}
For a given coding task, we combine demonstrations, queries, and natural language instructions using templates to construct prompts. Building on prior work~\cite{allyouneed,UBC}, we adapted these templates to suit our specific tasks, as outlined in Table~\ref{temp}. For instance, the instruction ``Generate only single line code'' is designed to ensure the LLM produces concise code that closely aligns with the reference solution, facilitating evaluation and comparison. These RAG-enhanced prompts supply the LLM with additional context regarding the developer's intent, which may improve the quality and relevance of the generated output.

\subsection{Implementation details}

\textbf{Large Language Model.} To ensure transparency and reproducibility of our experiments, we use one of the more recent open-source LLMs fine-tuned for code, CodeLlama-Instruct-13B~\cite{codellama} for all tasks. CodeLlama is trained based on Llama 2 for coding tasks, which supports large input contexts as well as zero-shot instructions. We set the temperature to 0 to get a more consistent output. Due to memory constraints, the maximum number of tokens is set to 7000. Thus, in the Assertion Generation task, we tested 1,444 samples in the Atlas test set that satisfy the length limit.



\textbf{Computing hardware.} We conduct all the experiments on a Linux server with four Nvidia RTX 3090 GPUs, AMD Ryzen 48-Core CPU, and 256 GB RAM.

\subsection{Approach of RQs}
\textbf{RQ1:} Our goal is to evaluate the effectiveness of various retrievers in the context of RAG for coding tasks. We assess their effectiveness by integrating them into the RAG process and comparing the generated outputs using our chosen evaluation metrics. A retriever is deemed more effective if its integration into RAG leads to better results.

To further understand why some retrievers outperform others, we examine which retrievers return better demonstrations during the retrieval phase. Unlike previous works~\cite{chronicle,bertbeyond}, coding tasks do not have explicitly defined \emph{ground truth} for relevant demonstrations. 
However, as shown in prior studies~\cite{oracle,whatmakes}, demonstrations that closely match the output sequences of a query, rather than its input sequences, usually can be referred to as \emph{gold demonstrations}, and provide the greatest benefit in improving the quality of the generated output. Notably, the output of the query is only accessible during the experiment. By using \emph{gold demonstrations} as the \emph{ground truth}, we can evaluate how well a retriever performs in retrieving demonstrations. To quantify this, we introduce a modified recall metric, \emph{RecallGold@K}, which is defined as the fraction of correctly retrieved \emph{gold demonstrations} from the top $K$ retrieved demonstrations. The formula is given by: 
\begin{equation}
\begin{aligned}
\emph{RecallGold@K} = \frac{\left|D_K \cap D_K^{Gold}\right|}{K}
\end{aligned}
\end{equation}

To mitigate the noise introduced by multiple demonstrations~\cite{longllmlingua,allyouneed}, we set $K$=1 in RQ1-4 and discuss the multiple-shot context in RQ5.

\textbf{RQ2:} The efficiency of a retriever is determined by \emph{how much time it takes to return the demonstrations for a query}. To evaluate this, we compared their \textbf{Time Per Query}~\cite{annbenchmarks}, aiming to identify the most efficient retrievers. Additionally, given the need for \textbf{scalability} in RAG when increasing the knowledge base, we examined how each retriever scales with the size of the knowledge base. This involved assessing retrieval efficiency under varying data loads.

\textbf{RQ3:} When deploying RAG systems, achieving a balance between efficiency and effectiveness is essential \cite{chronicle}. This requires considering both how quickly the retriever returns demonstrations and the quality of the final generated results. To observe this trade-off, we calculated the \textbf{Speedup} of other retriever relative to the state-of-the-art BM25 and the decrease in generation quality.  We also plot \textbf{Queries Per Second (QPS)} against the used evaluation metrics for measuring effectiveness, aiming to identify the optimal balance, typically located in the upper-right region of the plots.

\textbf{RQ4:} 
Approximate retrievers (LSH, ANNOY, HNSW) allow for tuning the degrees of approximation by adjusting hyper-parameter settings, which potentially affects their efficiency and effectiveness. In RQ1-3, we use the default hyper-parameter settings for the approximate dense retrievers. To identify the hyper-parameters that best balance effectiveness and efficiency, we conduct experiments with various configurations, including: 1) the number of nodes inspected for ANNOY, 2) the number of bits for LSH, and 3) the number of neighbors per graph for HNSW. Table~\ref{para} presents the hyper-parameters we investigated for each task.

\begin{table}[]
\caption{Hyper-parameter settings of approximate retrievers for each task.} 
\label{para}
\begin{tabular}{l|c|c|c}
\hline
Parameter                                                  & Program Synthesis & Commit    & Assertion  \\ \hline
\begin{tabular}[c]{@{}l@{}}ANNOY:\\ Search\_K\end{tabular} & {[}1,2,5,15,30{]} & {[}1,10,40,60,100{]} & {[}1,10,50,100,300{]} \\ \hline
\begin{tabular}[c]{@{}l@{}}LSH:\\ Nbits\end{tabular}       & {[}1,2,5,15,30{]} & {[}1,2,5,15,30{]}    & {[}1,2,5,15,40{]}     \\ \hline
\begin{tabular}[c]{@{}l@{}}HNSW:\\ M\end{tabular}          & {[}2,3,4,8,16{]}  & {[}2,3,4,8,32{]}     & {[}3,4,8,16,64{]}     \\ \hline
\end{tabular}
\end{table}

To provide additional insights, we also examine the retrieval \emph{Precision} of the approximate dense retrievers. \emph{Precision} represents the fraction of demonstrations in the result tuple that are true nearest neighbors, as determined through exhaustive dense retrieval without approximation~\cite{annbenchmarks}. It is given by:
\begin{equation}
\begin{aligned}
\emph{Precision@K} = \frac{\left|D_K^{Approx.}\cap D_K^{Exh.}\right|}{K}
\end{aligned}
\end{equation}

We also set $K$=1 here to avoid noise. Generally, tuning hyper-parameters to achieve higher degrees of approximate results in improved efficiency but lower retrieval precision. By analyzing this trade-off, we can assess how the hyper-parameters of approximate retrievers influence final generation effectiveness through their impact on retrieval precision.



\textbf{RQ5:} 
In previous work~\cite{howmany,whatmakes,longllmlingua}, it has been demonstrated that the number of demonstrations significantly impacts the effectiveness. However, they have examined the effect of shot count with single retrievers. To obtain more comprehensive insights, we conduct an investigation into how varying the number of demonstrations across different retrievers influences the quality of generated results. This systematic exploration aimed to determine the optimal number of shots across retrievers to maximize the effectiveness of the RAG.


\begin{table*}[htbp]
\centering
\caption{The results of retrieval approaches in terms of generation effectiveness and retrieval efficacy. A one-shot setting is employed to ensure a fair comparison and darker colors indicate superior results.(a) Program Synthesis ~(b) Commit Generation ~(c) Assertion Generation}
\label{tab:maintab}
\begin{tabular}{llccccccc}

            \\ \hline
\multicolumn{2}{c|}{(a) Approach}                                                                                         & Code Bleu                                          & Syntax Match                                        & Dataflow Match                                      & Ngram Match                                         & \begin{tabular}[c]{@{}c@{}}Time Per\\  Query (ms)\end{tabular}

& \multicolumn{1}{r}{Speedup} & \begin{tabular}[c]{@{}c@{}}Reduced \\ Code Bleu\end{tabular}                  \\ \hline
\multicolumn{1}{l|}{}                          & \multicolumn{1}{l|}{BM25L \cite{bm25L}}            & \cellcolor[HTML]{FFF2F2}0.249 & \cellcolor[HTML]{FFF2F2}0.379  & \cellcolor[HTML]{FFF2F2}0.361  & \cellcolor[HTML]{FFCCCC}0.128 & \cellcolor[HTML]{F86B6D}2.31 & 0.98x                           & 5.75\%$\downarrow$                                  \\
\multicolumn{1}{l|}{\multirow{-2}{*}{Sparse}}  & \multicolumn{1}{l|}{BM25 \cite{bm25}}              & \cellcolor[HTML]{FF6666}0.264 & \cellcolor[HTML]{FF6666}0.391  & \cellcolor[HTML]{FFCCCC}0.373  & \cellcolor[HTML]{FF6666}0.146 & \cellcolor[HTML]{F8696B}2.26 & 1.00x                            & 0.00\%-                                 \\ \hline
\multicolumn{1}{l|}{Exhaustive}                      & \multicolumn{1}{l|}{SBERT\_SS \cite{sentencebert}} & \cellcolor[HTML]{FF6666}0.261 & \cellcolor[HTML]{FF6666}0.387  & \cellcolor[HTML]{FFCCCC}0.380  & \cellcolor[HTML]{FF6666}0.138 & \cellcolor[HTML]{FFF2F2}5.07 & 0.45x                       & 1.21\%$\downarrow$                             \\ \hline
\multicolumn{1}{l|}{}                          & \multicolumn{1}{l|}{ANNOY \cite{annoy}}            & \cellcolor[HTML]{FFCCCC}0.259 & \cellcolor[HTML]{FFCCCC}0.380  & \cellcolor[HTML]{FF6666}0.385  & \cellcolor[HTML]{FFCCCC}0.136 & \cellcolor[HTML]{FFCCCC}5.29 & 0.43x    & 1.89{}\%$\downarrow$ \\
\multicolumn{1}{l|}{}                          & \multicolumn{1}{l|}{LSH \cite{faiss}}              & \cellcolor[HTML]{FFCCCC}0.252 & \cellcolor[HTML]{FFCCCC}0.385  & \cellcolor[HTML]{FFCCCC}0.376  & \cellcolor[HTML]{FFF2F2}0.124 & \cellcolor[HTML]{FFCCCC}5.08 & 0.44x    & 4.62{}\%$\downarrow$ \\
\multicolumn{1}{l|}{\multirow{-3}{*}{Approximate}} & \multicolumn{1}{l|}{HNSW \cite{hnsworg}}           & \cellcolor[HTML]{FFCCCC}0.258 & \cellcolor[HTML]{FFCCCC}0.384  & \cellcolor[HTML]{FF6666}0.381  & \cellcolor[HTML]{FFCCCC}0.134 & \cellcolor[HTML]{FFCCCC}5.18 & 0.44x    & 2.27{}\%$\downarrow$ \\ \hline
\multicolumn{2}{c|}{(b) Approach}                                                                                         & RougeL                                             & Meteor                                              & Rouge1                                              & Rouge2                                             &\begin{tabular}[c]{@{}c@{}}Time Per\\  Query (ms)\end{tabular}          & Speedup                    & \begin{tabular}[c]{@{}c@{}}Reduced \\ RougeL\end{tabular}                       \\ \hline
\multicolumn{1}{l|}{}                          & \multicolumn{1}{l|}{BM25L}                                          & \cellcolor[HTML]{F2FFF2}0.186 & \cellcolor[HTML]{F2FFF2}0.149  & \cellcolor[HTML]{F2FFF2}0.192  & \cellcolor[HTML]{F2FFF2}0.087 &  \cellcolor[HTML]{F2FFF2}252.98                       & 0.95x                           &35.34\%$\downarrow$                                 \\
\multicolumn{1}{l|}{\multirow{-2}{*}{Sparse}}  & \multicolumn{1}{l|}{BM25}                                           & \cellcolor[HTML]{66FF66}0.288 & \cellcolor[HTML]{66FF66}0.244  & \cellcolor[HTML]{66FF66}0.294  & \cellcolor[HTML]{66FF66}0.169 &  \cellcolor[HTML]{F2FFF2}239.53                       & 1.00x                           & 0.00\%-                                  \\ \hline
\multicolumn{1}{l|}{Exhaustive}                      & \multicolumn{1}{l|}{SBERT\_SS}                                      & \cellcolor[HTML]{66FF66}0.286 & \cellcolor[HTML]{CCFFCC}0.238  & \cellcolor[HTML]{CCFFCC}0.290  & \cellcolor[HTML]{CCFFCC}0.166 & \cellcolor[HTML]{CCFFCC}16.67                        & 14.37x                       & 1.11\%$\downarrow$                             \\ \hline
\multicolumn{1}{l|}{}                          & \multicolumn{1}{l|}{ANNOY}                                          & \cellcolor[HTML]{CCFFCC}0.280 & \cellcolor[HTML]{CCFFCC}0.234 & \cellcolor[HTML]{CCFFCC}0.286  & \cellcolor[HTML]{CCFFCC}0.164 & \cellcolor[HTML]{66FF66}5.83                         & 41.08x                       & 2.74{}\%$\downarrow$ \\
\multicolumn{1}{l|}{}                          & \multicolumn{1}{l|}{LSH}                                            & \cellcolor[HTML]{CCFFCC}0.285 & \cellcolor[HTML]{66FF66}0.242 & \cellcolor[HTML]{66FF66}0.291  & \cellcolor[HTML]{66FF66}0.168 & \cellcolor[HTML]{66FF66}5.82                         & 41.17x                       & 0.88{}\%$\downarrow$   \\
\multicolumn{1}{l|}{\multirow{-3}{*}{Approximate}} & \multicolumn{1}{l|}{HNSW}                                           & \cellcolor[HTML]{CCFFCC}0.283 & \cellcolor[HTML]{CCFFCC}0.237 & \cellcolor[HTML]{CCFFCC}0.289 & \cellcolor[HTML]{CCFFCC}0.166 & \cellcolor[HTML]{66FF66}5.44                         & 44.05x                       & 1.74{}\%$\downarrow$ \\ \hline
\multicolumn{2}{c|}{(c) Approach}                                                                                         & Exact Match                                        & \begin{tabular}[c]{@{}c@{}}Longest Common\\  Subsequence\end{tabular}                                & Plausible Match                                     & Edit Distance                                      & \begin{tabular}[c]{@{}c@{}}Time Per\\  Query (ms)\end{tabular}          & Speedup                     & \begin{tabular}[c]{@{}c@{}}Reduced \\Exact Match\end{tabular}                \\ \hline
\multicolumn{1}{l|}{}                          & \multicolumn{1}{l|}{BM25L}                                          & \cellcolor[HTML]{F2F2FF}0.414 & \cellcolor[HTML]{F2F2FF}73.3   & \cellcolor[HTML]{F2F2FF}0.492  & \cellcolor[HTML]{F2F2FF}15.9  & \cellcolor[HTML]{F2F2FF}206.19                       & 1.15x                           & 32.22\%$\downarrow$                                  \\
\multicolumn{1}{l|}{\multirow{-2}{*}{Sparse}}  & \multicolumn{1}{l|}{BM25}                                           & \cellcolor[HTML]{6666FF}0.612 & \cellcolor[HTML]{6666FF}83.6   & \cellcolor[HTML]{6666FF}0.671  & \cellcolor[HTML]{6666FF}7.5   & \cellcolor[HTML]{F2F2FF}237.53                       & 1.00x                           & 0.00\%-                                  \\ \hline
\multicolumn{1}{l|}{Exhaustive}                      & \multicolumn{1}{l|}{SBERT\_SS}                                      & \cellcolor[HTML]{6666FF}0.603 & \cellcolor[HTML]{6666FF}83.2   & \cellcolor[HTML]{6666FF}0.669  & \cellcolor[HTML]{CCCCFF}8.1   & \cellcolor[HTML]{CCCCFF}97.09                        & 2.45x                       & 1.32\%$\downarrow$                             \\ \hline
\multicolumn{1}{l|}{}                          & \multicolumn{1}{l|}{ANNOY}                                          & \cellcolor[HTML]{CCCCFF}0.576 & \cellcolor[HTML]{CCCCFF}81.8   & \cellcolor[HTML]{CCCCFF}0.649  & \cellcolor[HTML]{CCCCFF}8.9   & \cellcolor[HTML]{6666FF}6.34                         & 37.45x                     & 5.74\%$\downarrow$ \\
\multicolumn{1}{l|}{}                          & \multicolumn{1}{l|}{LSH}                                            & \cellcolor[HTML]{CCCCFF}0.601 & \cellcolor[HTML]{CCCCFF}83.0   & \cellcolor[HTML]{CCCCFF}0.666  & \cellcolor[HTML]{6666FF}8.0   & \cellcolor[HTML]{6666FF}7.79                         & 30.51x                       & 1.78{}\%$\downarrow$ \\
\multicolumn{1}{l|}{\multirow{-3}{*}{Approximate}} & \multicolumn{1}{l|}{HNSW}                                           & \cellcolor[HTML]{CCCCFF}0.597 & \cellcolor[HTML]{CCCCFF}82.7   & \cellcolor[HTML]{CCCCFF}0.664  & \cellcolor[HTML]{CCCCFF}8.3   & \cellcolor[HTML]{6666FF}6.06                         & 39.21x                      & 2.35{}\%$\downarrow$ \\ \hline
\end{tabular}
\end{table*}

\section{RESULTS}





\begin{figure}[tbp]
\centering
\subfigure[Program Synthesis ~(b) Commit Generation ~ (c) Assertion Generation]
{\includegraphics[width=\linewidth]{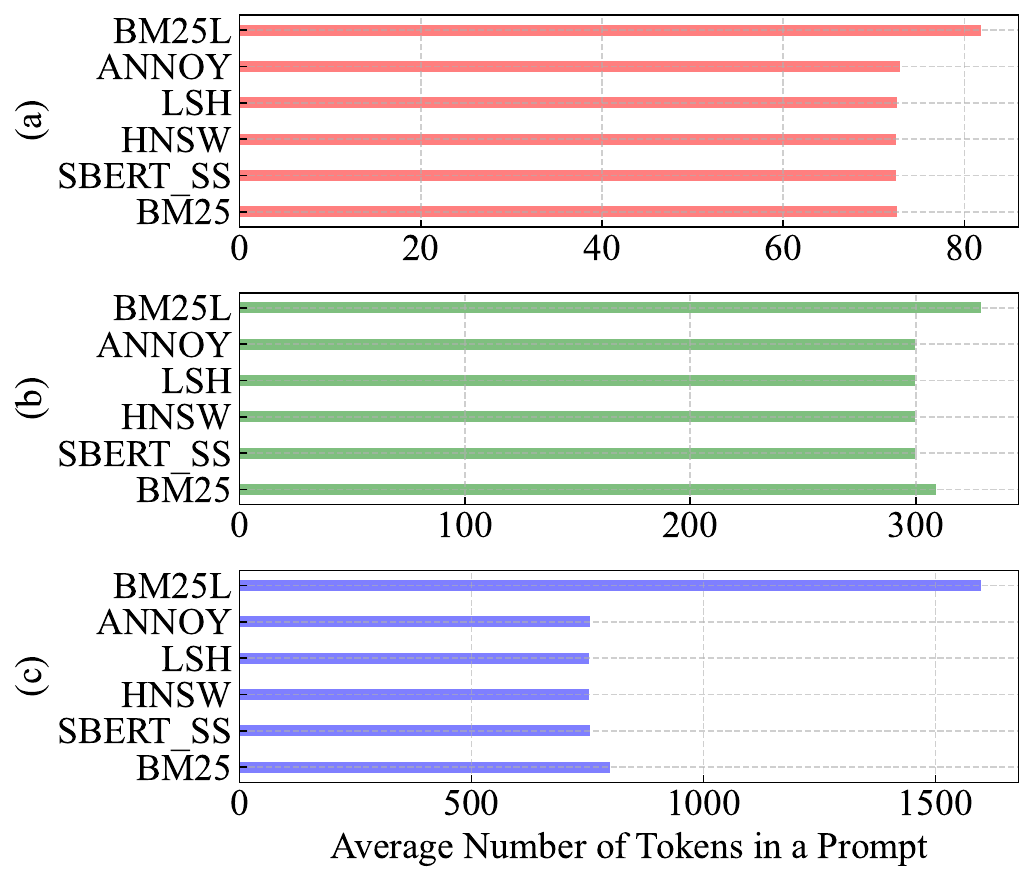}}
\caption{The average tokens of prompts of different retrievers.}
\label{fig:length}
\end{figure}

\subsection{RQ1 - Effectiveness}\label{sec:rq1}
Table~\ref{tab:maintab} presents the empirical results of generated outputs across three coding tasks. For example, BM25 achieves a Code Bleu rate of 0.264 in the Program Synthesis task, a RougeL score of 0.288 in the Commit Generation task, and an Exact Match score of 0.612 in the Assertion Generation task. Notably, BM25 outperforms all other retrievers across these tasks. However, despite being a sparse retriever, BM25L consistently performs worse compared to other retrievers.

The underperformance of BM25L may be attributed to the length preference bias it introduces. Specifically, BM25L modifies the term frequency normalization to favor longer documents\cite{bm25L}. As illustrated in Fig.~\ref{fig:length}, the demonstration returned by BM25L consists of longest prompts. This inclination may cause the BM25L to select longer rather than relevant demonstrations, compromising its effectiveness.

When comparing effectiveness within dense retrievers, we observe that the exhaustive dense retriever SBERT\_SS generally performs better than the three approximate dense retrievers, though the gap is always small. This indicates that while SBERT\_SS achieves slightly higher effectiveness, the approximate dense retrievers offer a comparable level of effectiveness while potentially offering gains in efficiency.
\begin{figure}[tbp]
\centering
\subfigure[Program Synthesis ~(b) Commit Generation ~ (c) Assertion Generation]
{\includegraphics[width=\linewidth]{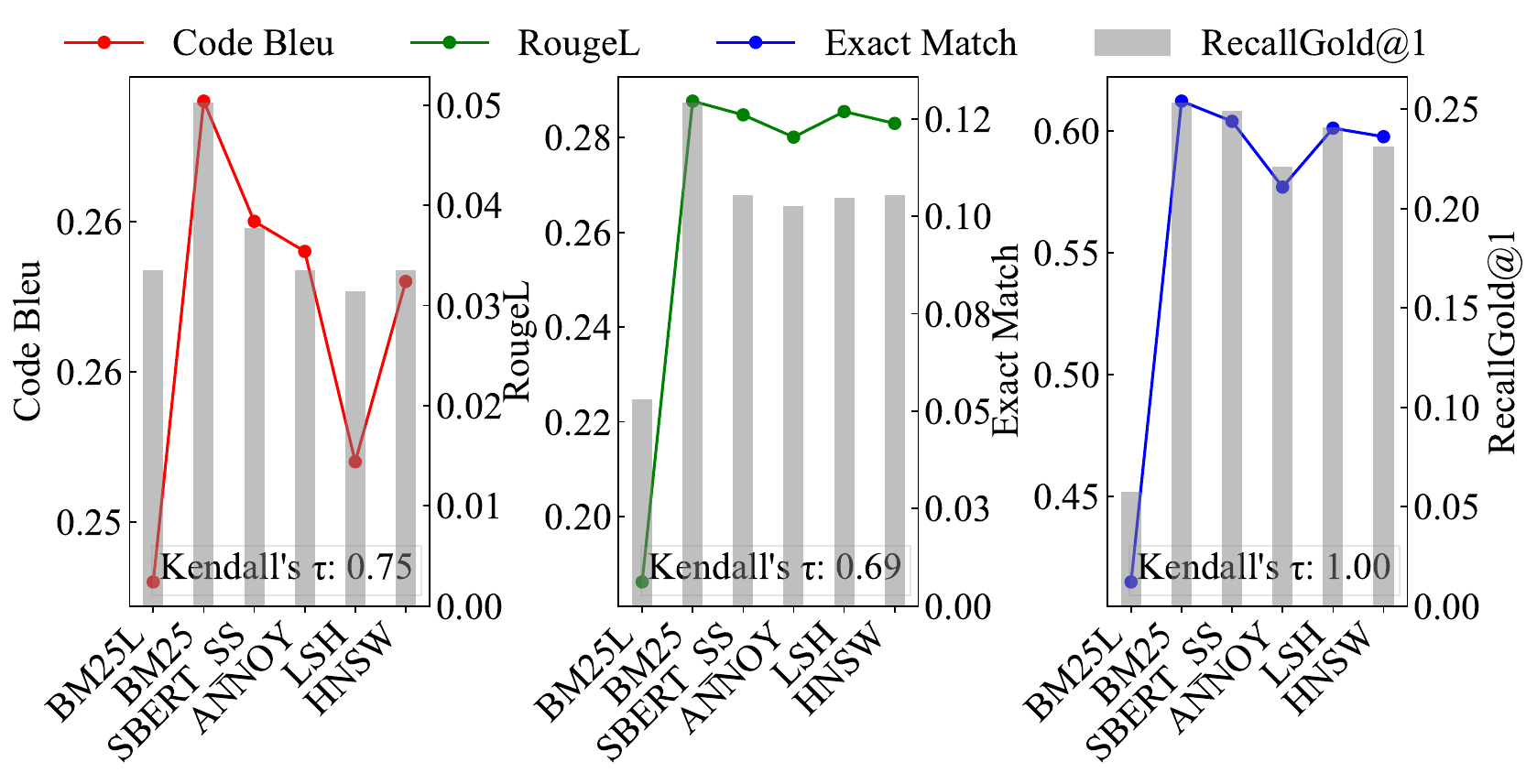}}
\caption{The effectiveness of retrievers exhibits a strong correlation to the retriever's ability to recall \emph{gold demonstrations}.}
\label{fig:recallgold}
\end{figure}

To understand why some retrievers outperform others, we examine which retrievers return better demonstrations using \emph{RecallGold@K}. As illustrated in Fig.~\ref{fig:recallgold}, we observe a strong positive correlation between the \emph{RecallGold@K} scores of various retrievers and their effectiveness. The non-parametric Kendall's $\tau$ correlations \cite{kendall} are 0.75, 0.69, and 1.00 for Program Synthesis, Commit Generation, and Assertion Generation, respectively, with corresponding p-values of 0.04, 0.05, and 0.003. These findings suggest that the effectiveness of retrievers is significantly influenced by their ability to recall \emph{gold demonstrations}.

\begin{mdframed}[roundcorner=5pt]
\noindent
\textbf{Finding 1:} The effectiveness hierarchy of retrievers is BM25 \textgreater\xspace exhaustive dense (SBERT\_SS) $\gtrsim$ approximate dense (ANNOY, LSH, HNSW) \textgreater\xspace BM25L. The effectiveness of retrievers exhibits a strong correlation to the retriever's ability to recall \emph{gold demonstrations}.
\end{mdframed}

\subsection{RQ2 - Efficiency}\label{sec:rq2}

\begin{figure}[]

\centering
\subfigure[Program Synthesis ~(b) Commit Generation ~ (c) Assertion Generation]
{\includegraphics[width=\linewidth]{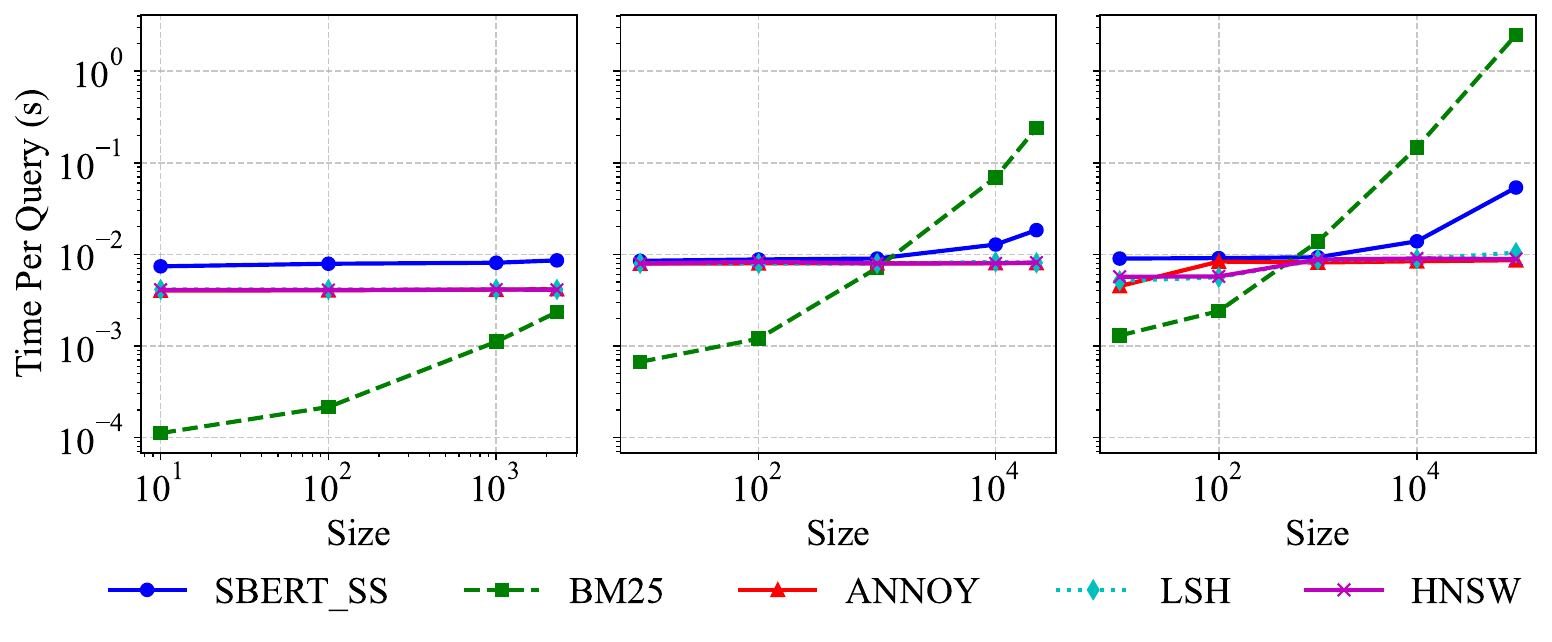}}
\caption{Efficacy comparison between different retrievers with increasing knowledge base size.}
\label{fig:kbsize}
\end{figure}

We present the efficiency of various retrievers across the three coding tasks in Table~\ref{tab:maintab}. For instance, in the Program Synthesis task, it takes 2.26 ms for BM25 to retrieve demonstrations for a query. In this task, the sparse retrievers tend to have the shortest retrieval times, while the dense retrievers take longer. However, this trend reverses in the Commit Generation and Assertion Generation tasks, where dense retrievers outperform sparse retrievers in terms of efficiency. 

Among the dense retrievers, there is minimal efficiency disparity when the knowledge base is small, as observed in the Program Synthesis task. However, as the knowledge base grows, approximate dense retrievers demonstrate significant efficiency advantages. For instance, in the Assertion Generation task, ANNOY reduces retrieval time to 6 ms compared to the 97 ms required by the exhaustive SBERT\_SS retriever. This difference arises because SBERT\_SS performs a linear scan of all demonstrations in the knowledge base, which can be computationally intensive. In contrast, approximate dense retrievers like HNSW, ANNOY, and LSH leverage specialized data structures that allow for a degree of approximation, significantly reducing the number of comparisons needed. The variation in efficiency across tasks suggests that the size of the knowledge base plays a crucial role in influencing the efficiency of different retrievers. 

In addition to analyzing fixed-size knowledge base, we also investigate scalability in terms of efficiency. Fig. \ref{fig:kbsize} illustrates the efficiency trends of each retriever as the knowledge base size increases. To simplify the comparison, we only plot BM25, as it exhibits similar efficiency to its variant, BM25L. At smaller knowledge base sizes (fewer than $10^3$ entries), BM25 shows greater efficiency. However, as the knowledge base expands, its efficiency decreases relative to that of dense retrievers. Dense retrievers, which utilize fixed-size numeric representations, become increasingly efficient compared to sparse retrievers that need to manage large document-term matrices. Furthermore, approximate dense retrievers such as HNSW and ANNOY exhibit significant scalability, maintaining a nearly flat efficiency curve as the knowledge base size increases. This demonstrates the advantages of using approximate dense retrievers for large-scale retrieval tasks.


\begin{mdframed}[roundcorner=5pt]
\noindent
\textbf{Finding 2:} For knowledge bases smaller than $10^3$, the efficiency hierarchy is sparse \textgreater\xspace exhaustive dense \textgreater\xspace approximate dense. In more general cases with knowledge bases larger than 
$10^3$, the efficiency hierarchy changes to approximate dense \textgreater\xspace exhaustive dense \textgreater\xspace sparse.
\end{mdframed}

\subsection{RQ3 - Trade-off between Efficiency and Effectiveness}\label{sec:rq3}

\begin{figure*}[]
\centering
\subfigure[Program Synthesis ~(b) Commit Message Generation ~ (c) Assertion Generation]{\includegraphics[width=\linewidth]{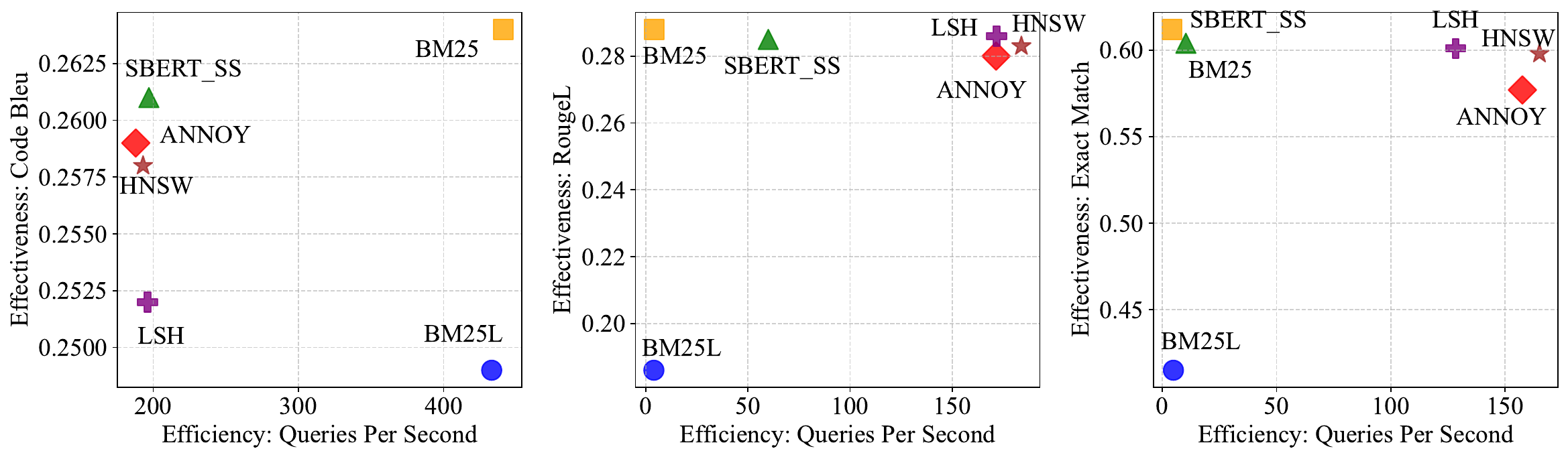}}
\caption{The trade-off between efficiency and effectiveness of retrievers. The retrievers located in up and right corner are better.}
\label{fig:tradeoff}
\end{figure*}

Based on the previous RQs, BM25 emerged as the most effective but less efficient retriever for tasks involving large knowledge bases. Using BM25 as a benchmark, we presented the speedup and quality degradation of other retrievers in Table \ref{tab:maintab}. For instance, in Assertion Generation tasks, while exhaustive dense retrieval offers a 2x speedup with only a 1.32\% drop in Exact Match, LSH provides a further 31x speedup with just a 1.78\% reduction through approximation. Similarly, in Commit Generation, HNSW achieves a 44x speedup, while only with a 1.74\% drop in RougeL compared with BM25.

We also plot both efficiency and effectiveness of different retrievers in Fig.~\ref{fig:tradeoff} to provide a trade-off perspective. Retrievers positioned in the upper-right corner represent a more favorable balance between efficiency and effectiveness.

This trade-off varies across tasks. In the Program Synthesis task, BM25 achieves the optimal balance, combining high effectiveness with strong efficiency. In contrast, SBERT\_SS and BM25L fall short: SBERT\_SS is located in the upper-left quadrant, indicating high effectiveness but low efficiency, while BM25L resides in the lower-right quadrant, highlighting poor effectiveness and decent efficiency. Additionally, the three approximate dense retrievers (ANNOY, LSH, and HNSW) underperform in both dimensions in this task. 
In the Commit Generation and Assertion Generation tasks, BM25 and SBERT\_SS start with high effectiveness but exhibit low efficiency, occupying the upper-left quadrant. BM25L continues to lag behind in both metrics. However, all three approximate dense retrievers, ANNOY, LSH, and HNSW, strike a much better balance between efficiency and effectiveness, outperforming their sparse and exhaustive dense counterparts in these tasks with a larger knowledge base.

\begin{mdframed}[roundcorner=5pt]
\noindent
\textbf{Finding 3:} BM25 achieves the best balance between efficiency and effectiveness when the knowledge base size is small. However, for tasks involving larger knowledge bases, approximate dense retrievers offer the best trade-off between efficiency and effectiveness.
\end{mdframed}

\subsection{RQ4 - Impact of the Hyper-parameters}
\begin{figure}[tbp]
\centering

\subfigure[Program Synthesis ~(b) Commit Generation ~ (c) Assertion Generation]{\includegraphics[width=\linewidth]{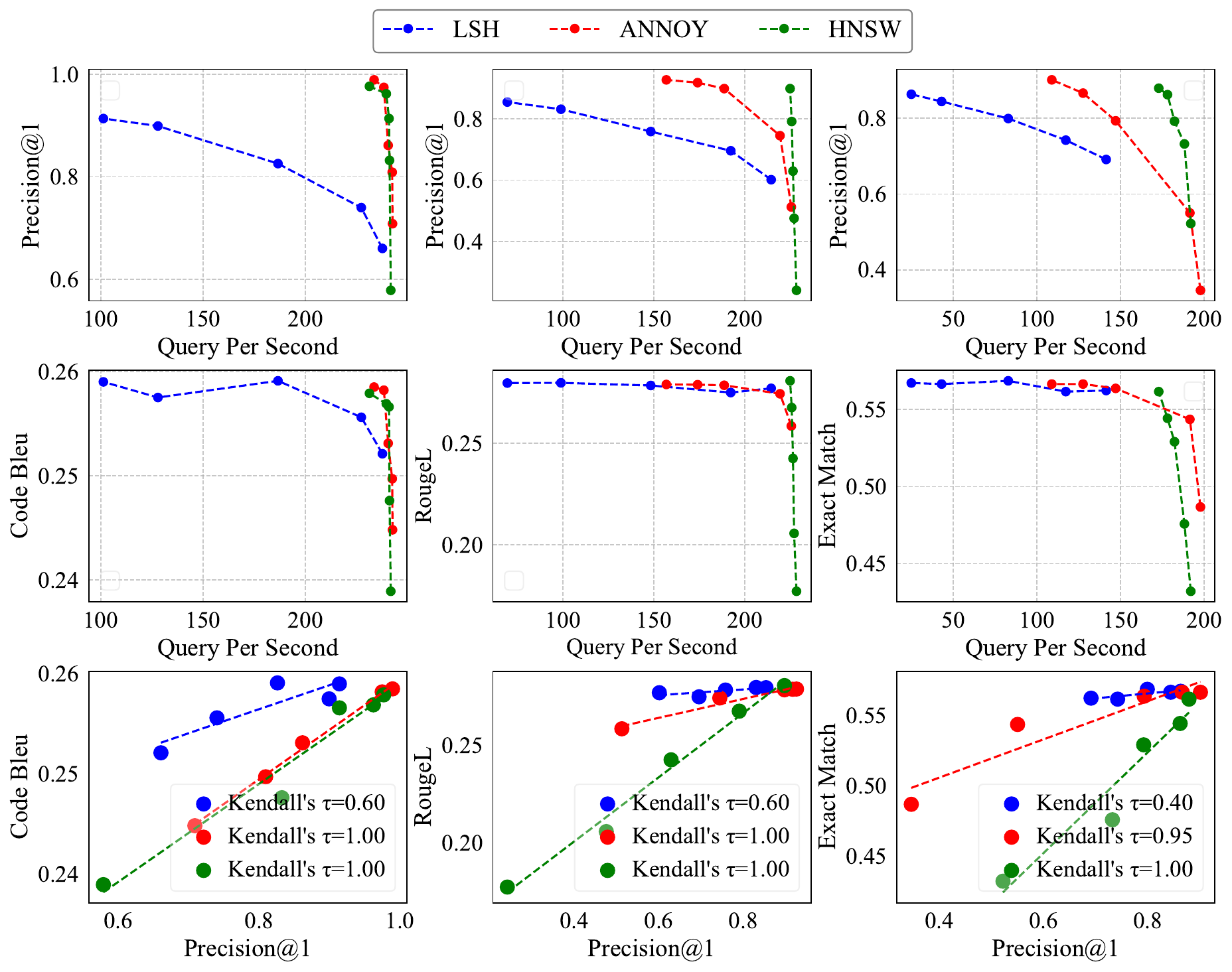}}
\caption{Effectiveness-efficiency (top), precision-efficiency (middle), and precision-effectiveness (bottom) relations for approximate retrievers with different hyper-parameters.}
\label{fig:9}
\end{figure}

Fig.~\ref{fig:9} illustrates the relationships between effectiveness and efficiency (top row), retrieval precision and efficiency (middle row), retrieval precision and effectiveness (bottom row) using different hyper-parameter configurations. A high degree of approximation (i.e., a smaller hyper-parameter setting in Table~\ref{para}) always leads to higher efficiency (higher QPS), but lower retrieval precision and effectiveness since ignores more nodes inspected. Specifically, LSH consistently maintains stable effectiveness and precision across different QPS values, with only slight variations. HNSW achieves high QPS values across a range of parameters. However, at the highest QPS values, the retrieval precision and effectiveness of generated outputs drops significantly. ANNOY demonstrates a similar trade-off to HNSW, though the decline in precision and effectiveness is less pronounced. 

Fig.~\ref{fig:9} also shows that the trade-off between retrieval precision and effectiveness remain consistent due to their strong positive correlation. Both HNSW and ANNOY's Kendall $\tau$ coefficients were statistically significant (p-value $\leq$ 0.05). In contrast, LSH exhibited p-values of 0.23, 0.23, and 0.48 across the three tasks, potentially due to the low variance of LSH's effectiveness metrics. Therefore, optimizing these trade-off during the retrieval step is essential to achieve the optimal balance between efficiency and effectiveness in the final generation phase.
\begin{mdframed}[roundcorner=5pt]
\noindent
\textbf{Finding 4:}  A higher degree of approximation improves the efficiency of dense retrieval but reduces retrieval precision, which adversely affects overall effectiveness. This is because of the strong positive correlation between retrieval precision and effectiveness.
\end{mdframed}



\subsection{RQ5 - Impact of Number of Shots}\label{sec:rq5}

\begin{figure}[bp]
\centering
\subfigure[Program Synthesis ~(b) Commit Generation ~ (c) Assertion Generation]
{\includegraphics[width=\linewidth]{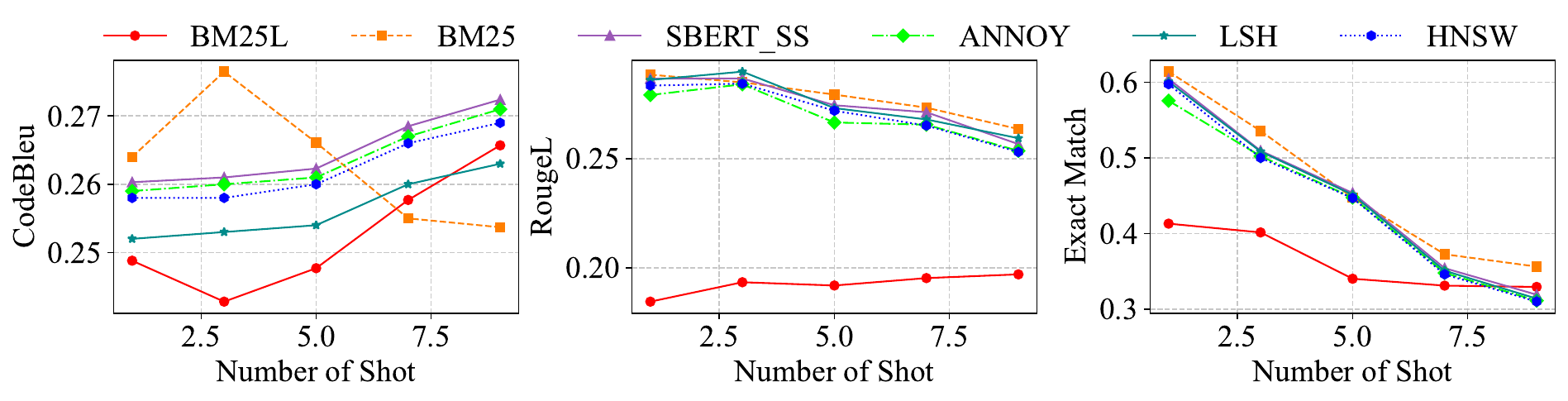}}
\caption{The impact of the number of shots on the effectiveness of different retrievers.}
\label{fig:shot}
\end{figure}

\begin{table}[ht]
\centering
\caption{Case study from Program Synthesis: LLMs can be easily distracted by noisy information.}
\label{demonstration}
\begin{tabular}{l|p{1cm}|p{5.5cm}}
\hline
\multirow{2}{*}{Sample} & Query & \begin{tabular}[c]{@{}p{7cm}@{}} \#\#\# requirement\\ create list of values from dictionary `dict1` that \\ have a key that starts with 'EMP' \\ \#\#\# code \end{tabular} \\ \cline{2-3}
                        & Actual & {[}value for key, value in list(dict1.items()) if key.startswith('EMP'){]} \\ \hline
\multirow{2}{*}{1-shot} & Context & \begin{tabular}[c]{@{}p{7cm}@{}} Demo1: \#\#\# requirement \\ create a list of values from the dictionary \\ `programs` that have a key with a case \\ insensitive match to 'new york' \\ \#\#\# code \\ {[}value for key, value in list(programs.items()) if\\ 'new york' in key.lower(){]} \end{tabular} \\ \cline{2-3}
                        & Response & {[}value for key, value in list(dict1.items()) if key.startswith('EMP'){]} \textcolor{green}{\ding{51}} \\ \hline
\multirow{2}{*}{2-shot} & Context & \begin{tabular}[c]{@{}p{7cm}@{}} //Demo1 \\ Demo2: \#\#\# requirement \\ sort a dictionary `a` by values that are list type \\ \#\#\# code \\ t = sorted(list(a.items()), key=lambda x: x{[}1{]}) \end{tabular} \\ \cline{2-3}
                        & Response & {[}value for key, value in list(dict1.items()) if key.startswith('EMP'){]} \textcolor{green}{\ding{51}} \\ \hline
\multirow{2}{*}{3-shot} & Context & \begin{tabular}[c]{@{}p{7cm}@{}} //Demo1 \\ //Demo2 \\ Demo3: \#\#\# requirement \\ get keys of dictionary `my\_dict` that contain any\\ values from list `lst` \\ \#\#\# code \\ {[}key for key, value in list(my\_dict.items()) if \\set(value).intersection(lst){]} \end{tabular} \\ \cline{2-3}
                        & Response & {[}key for key, value in list(dict1.items()) if key.startswith('EMP'){]} \textcolor{red}{\ding{55}} \\ \hline
\end{tabular}
\end{table}

Fig.~\ref{fig:shot} illustrates the relationship between effectiveness and the number of demonstrations across different retrievers. Notably, more demonstrations do not always correlate with improved effectiveness. In Assertion Generation, the Exact Match rate consistently decreases as the number of demonstrations increases across all retrievers. In Program Synthesis, while three demonstrations yield better results than one when using BM25, adding more than three leads to a decline in effectiveness. Specifically, we not only demonstrate degradation as observed in prior works~\cite{howmany,whatmakes,longllmlingua}, but also provide a detailed case study to illustrate why this degradation occurs.
A potential explanation for introducing more demonstrations does not assert improvement is that too many demonstrations may introduce noise, as LLMs can easily become distracted~\cite{distracte}. This effect is particularly evident when the inputs of multiple demonstrations are highly similar, but their outputs differ significantly. Table~\ref{demonstration} presents an example where the LLM's response to the same query changes as more demonstrations are added. Although the inputs in the query and demonstrations are relevant, the corresponding code examples address distinct objectives. For example, Demo3 introduces code for obtaining keys based on dictionary values, while the query requires retrieving values based on key prefixes. 

This also explains the inconsistent trends observed as the number of demonstrations increases across different tasks. As shown in Table~\ref{latency}, the average input similarity between the demonstrations in the retrieval tuple and the query is significantly higher in Assertion Generation and Commit Generation tasks compared to Program Synthesis. For instance, in Assertion Generation, demonstrations in 9-shot prompts exhibit a high average similarity of 0.819 to the query, yet their outputs are likely to differ. This increases the likelihood that the LLM will become distracted in this task. These observations align with the findings of Gao et al.~\cite{whatmakes}, which concluded that a more diverse set of demonstrations tends to be more beneficial for RAG, as it reduces the risk of confusing the LLM with highly similar demonstrations that have divergent outputs.

Table~\ref{latency} also shows that the inclusion of additional demonstrations, despite their uncertain benefits, inevitably increases the latency of LLM responses due to the longer prompt length. Therefore, the careful selection of the number of demonstrations is critical and should be tailored to the specific requirements of the task at hand.


\begin{table}[]
\caption{The average prompt length, query latency (ms), and cosine similarity (\%) of demonstrations to the query using SBERT\_SS with varying numbers of shots.}
\label{latency}
\begin{tabular}{l|ccccc}
\hline
\textbf{Task}               & \multicolumn{5}{c}{Program Synthesis}     \\ \hline
\textbf{\# of shots}        & 1     & 3      & 5      & 7      & 9      \\
\textbf{Avg \# of Tokens}   & 72.4  & 173    & 272.6  & 372    & 471.6  \\
\textbf{Latency Per Query}  & 1.8   & 1.9    & 2.3    & 2.5    & 2.6    \\
\textbf{Avg Similarity} & 72.9  & 69.8   & 67.8   & 66.4   & 65.2   \\ \hline
\textbf{Task}               & \multicolumn{5}{c}{Commit Generation}     \\ \hline
\textbf{\# of shots}        & 1     & 3      & 5      & 7      & 9      \\
\textbf{Avg \# of Tokens}   & 299.8 & 540.2  & 779.1  & 1017.2 & 1254.5 \\
\textbf{Latency Per Query}  & 0.5   & 0.6    & 0.7    & 0.9    & 1      \\
\textbf{Avg Similarity} & 88.9  & 86.4   & 85     & 84.2   & 83.5   \\ \hline
\textbf{Task}               & \multicolumn{5}{c}{Assertion Generation}  \\ \hline
\textbf{\# of shots}        & 1     & 3      & 5      & 7      & 9      \\
\textbf{Avg \# of Tokens}   & 755.1 & 1515.3 & 2263.3 & 3004.2 & 3747.7 \\
\textbf{Latency Per Query}  & 0.9   & 1.2    & 1.6    & 1.9    & 2.2    \\
\textbf{Avg Similarity} & 91.5  & 87.5   & 85.1   & 83.2   & 81.9   \\ \hline
\end{tabular}
\end{table}

\begin{mdframed}[roundcorner=5pt]
\noindent
\textbf{Finding 5:} More demonstrations are not always beneficial to RAG. Adding additional demonstrations may introduce more noise, potentially leading to inaccurate outputs. Moreover, an increase in the number of tokens in the prompt can result in longer response delays.
\end{mdframed}

\section{Discussion}\label{sec:discussion}

\subsection{Implications of our study}\label{sec:implications}

\textbf{We encourage practitioners to select retrievers based on their specific context. Generally, dense approximate retrievers offer the best balance between effectiveness and efficiency.} 
In light of Findings 1, 2, and 3, different retrievers vary in their efficiency and effectiveness. For instance, with knowledge bases smaller than $10^3$, the sparse retriever BM25 is more efficient and effective compared to approximate dense and exhaustive dense retrievers. However, for larger knowledge bases exceeding $10^3$, approximate dense retrievers demonstrate superior efficiency while maintaining effectiveness comparable to BM25. Therefore, considering both efficiency and effectiveness, employing an approximate dense retriever can enhance speed with only a minimal trade-off in generation quality, except when dealing with very small knowledge bases.

\textbf{We recommend practitioners balance effectiveness and efficiency when tuning hyper-parameters for approximate retrievers.}
As demonstrated in RQ4, the parameters of approximate dense retrievers can be adjusted to improve retrieval efficiency. However, practitioners should be cautious, as retrieval precision and overall effectiveness may significantly decline at high QPS (Queries Per Second). 

\textbf{More demonstrations do not necessarily guarantee better effectiveness. Practitioners should exercise caution when adding demonstrations.} 
As indicated in RQ5, increasing the number of demonstrations does not necessarily improve the effectiveness of RAG. In certain instances, adding more demonstrations can distract the model, particularly when the retrieved demonstrations share similar inputs but yield different outputs relative to the query. Consequently, it is advisable to limit the number of demonstrations.

\subsection{Threats to validity}\label{sec:threats}

\noindent\textbf{Internal Validity}
Prompt engineering has a significant impact on the LLM's performance~\cite{grabb2023impact}. Different prompts probably can lead to different results. However, as we discussed in Section~\ref{sec:experimentalsetting}, our tasks are basic and straightforward. More importantly, we use the same prompt template for our experiments for fair comparison.

\noindent\textbf{External Validity}
relates to the generalizability of our findings. In this study, we opted to use CodeLlama-13b for our experiments due to the cost constraints, RAG capability, and memory limitation. This choice may influence the generalizability of our findings.  Similarity, we focus on six retrievers in this study, which are selected from different families, and widely used in previous studies. Nevertheless, we encourage future study to extend our study to more LLMs and retrievers.

\section{Conclusion}
In this paper, we experimentally investigate the efficiency-effectiveness trade-off of different retrievers applied to RAG in coding tasks. Our results show that the sparse BM25 retriever excels in effectiveness, particularly in recalling more \emph{gold demonstrations}. However, the trade-off between efficiency and effectiveness depend on the size of the knowledge base. Sparse approaches are more efficient only when the knowledge base contains fewer than $10^3$ entries. For larger-scale retrieval, approximate dense retrievers offer significant efficiency gains with minimal degradation in effectiveness. Additionally, our findings reveal that increasing the number of demonstrations does not always enhance RAG effectiveness. In fact, it can introduce response latency and lead to incorrect outputs due to distraction. These insights offer valuable guidance for practitioners aiming to build efficient and effective RAG systems for coding tasks.

\bibliographystyle{IEEEtran}
\balance
\bibliography{0-main.bib}
\end{document}